\documentclass[graybox]{svmult}

\usepackage[]{latexsym,amssymb,amsmath}
\usepackage{mathptmx}       
\usepackage{helvet}         
\usepackage{courier}        
\usepackage{type1cm}        
\usepackage{makeidx}         
\usepackage{graphicx}        
\usepackage{multicol}        
\usepackage[bottom]{footmisc}
\usepackage{url}
\newcommand{\be}{\begin{eqnarray}}
\newcommand{\ee}{\end{eqnarray}}
\newcommand{\ud}{\mathrm{d}}

\newcommand{\gn}{G_{\rm N}}
\newcommand{\lp}{\ell_{\rm p}}
\newcommand{\mpl}{m_{\rm p}}

\makeindex             

\begin{document}

\title*{Quantum Harmonic Black Holes}
\author{Roberto Casadio, Alessio Orlandi}
\institute{Roberto Casadio \at Dipartimento di Fisica e Astronomia, Universit\`a di Bologna,
via Irnerio~46, 40126 Bologna, Italy \email{casadio@bo.infn.it}
\and Alessio Orlandi \at Dipartimento di Fisica e Astronomia, Universit\`a di Bologna,
via Irnerio~46, 40126 Bologna, Italy \email{alessio.j.orlandi@gmail.com}}
%
%
\maketitle

\abstract*{Inspired by the recent conjecture that black holes are condensates of gravitons,
we investigate a simple model for the black hole degrees of freedom that is consistent
both from the point of view of Quantum mechanics and of General Relativity.
Since the two perspectives should ``converge'' into a unified picture for small,
Planck size, objects, we expect our construction is a useful step for understanding
the physics of microscopic, quantum black holes.
In particular, we show that a harmonically trapped condensate gives rise to
two horizons, whereas the extremal case (corresponding to a remnant with
vanishing Hawking temperature) is not contained in the spectrum.}

\abstract{Inspired by the recent conjecture that black holes are condensates of gravitons,
we investigate a simple model for the black hole degrees of freedom that is consistent
both from the point of view of Quantum mechanics and of General Relativity.
Since the two perspectives should ``converge'' into a unified picture for small,
Planck size, objects, we expect our construction is a useful step for understanding
the physics of microscopic, quantum black holes.
In particular, we show that a harmonically trapped condensate gives rise to
two horizons, whereas the extremal case (corresponding to a remnant with
vanishing Hawking temperature) is not contained in the spectrum.}

\section{Introduction}
\label{sec:1}
One of the major mysteries in modern theoretical physics is to understand
what are the internal degrees of freedom of black holes (BHs).
Our best starting point is the
classical description of BHs provided by General Relativity~\cite{chandra92},
along with well established semiclassical results, such as the
predicted Hawking radiation~\cite{hawking74,hawking75}.
\par
It was recently proposed by Dvali and Gomez that BHs are
Bose-Einstein Condensates (BECs) of gravitons at a critical point,
with Bogoliubov modes that become degenerate and nearly gapless
representing the holographic quantum degrees of freedom responsible
for the BH entropy and the information storage~\cite{dvali12,dvali1207,dvali13,dvali13b}.
In order to support this view, they consider a collection of objects (gravitons)
interacting via Newtonian gravitational potential $ V_{\rm N}
\sim
-\frac{\gn\,\mu}{r}
\quad$ and whose effective mass $\mu$ is related to their characteristic quantum mechanical
size via the Compton/de~Broglie wavelength $\ell
\simeq
\frac{\hbar}{\mu}
=
\lp\,\frac{\mpl}{\mu}
$
.

These bosons can superpose and form a ``ball'' of radius $\ell$,
and total energy $M=N\,\mu$, where $N$ is the total number of constituents.
Within the Newtonian approximation, there is then a value of $N$
for which the whole system becomes a BH.
In details, given the coupling constant
$
\alpha
=
\frac{\lp^2}{\ell^2}
=
\frac{\mu^2}{\mpl^2}
$
there exists an integer $N$ such that no constituent can escape the 
gravitational well it contributed to create, and which can be approximately
described by the potential
\begin{equation}
 U(r)
\simeq
V_{\rm N}(\ell)
\simeq
-N\,\alpha\,\frac{ \hbar}{\ell}\,\Theta(\ell-r)
\ ,
\label{Udvali}
\end{equation}
where $\Theta$ is the Heaviside step function.
This implies that components in the depleting region
are ``marginally bound'' when the kinetic energy given by $E_K\simeq \mu$ equals the potential energy
\begin{equation}
E_K + U
\simeq
0
\quad
\Longleftrightarrow
N\,\alpha = 1
\label{energy0}
\ .
\end{equation}
Consequently, the effective boson mass and total mass of the
BH scale according
to
\begin{equation}
\mu
\simeq
\frac{\mpl}{\sqrt{N}}
\quad
{\rm and}
\quad
M
=
N\,\mu
\simeq
\sqrt{N}\,\mpl
\ .
\label{Max}
\end{equation}
Note that one has here assumed the ball is of size $\ell$
(since bosons superpose) and, therefore, the constituents will
interact at a maximum distance of order $r\sim\ell$, with fixed $\ell$.
The Hawking radiation and the negative specific heat
spontaneously result from quantum depletion of the condensate
for the states satisfying Eq.~\eqref{energy0}.
This description is partly Quantum Mechanics and partly classical Newtonian physics,
but no General Relativity is involved, in that geometry does not appear in the argument.
\par
%
%
%
\section{Quantum Mechanical Model}
We can improve on the former model by employing the Quantum Mechanical theory of the 
harmonic oscillator as a (better) mean field approximation for the Newtonian gravitational
interaction acting on each boson inside the BEC \footnote{We shall use units with $c=1$, $\hbar=\lp\,\mpl$ and the Newton constant
$\gn=\lp/\mpl$.}.
The potential $U$ in Eq.~\eqref{Udvali} is therefore replaced by~
\footnote{This is nothing but
Newton oscillator, which would correspond to a homogenous BEC distribution in the
Newtonian approximation.}
\begin{equation}
V = \frac{1}{2}\mu\omega^2(r^2-d^2)\Theta(d-r)
\equiv V_0(r)\Theta(d-r)
\label{Vho}
\end{equation}
and we further set $V(0)=U(0)$, so that $
 \frac{1}{2}\,\mu \, \omega^2 \, d^2 = N\,\alpha\,\frac{ \hbar}{\ell}
$.
We also assume that the effective mass, length and frequency of a single graviton mode
are related by $\mu = \hbar\, \omega = {\hbar}/{\ell}$, which leads to
$d =\sqrt{2\, N\, \alpha} \, \ell
=\sqrt{2\, N} \, \lp
$.
\par
If we neglect the finite size of the well, the Schr\"odinger equation in polar coordinates
yields the well-known eigenfunctions
\be
\psi_{nlm}(r,\theta,\phi)
=
\mathcal{N}\,r^l\, e^{-\frac{r^2}{ 2\, \ell^2}}\, _1F_1(-n,l+3/2,r^2/\ell^2)\, Y_{lm}(\theta,\phi)
\ ,
\label{psin}
\ee
where $\mathcal{N}$ is a normalization constant, $_1F_1$ the Kummer confluent
hypergeometric function of the first kind and $Y_{lm}(\theta,\phi)$
are the usual spherical harmonics.
The corresponding energy eigenvalues are given by
$
E_{nl}
=
\hbar\, \omega \left[2\,n+l + \frac{3}{2} - V(0) \right]
=
\hbar\,\omega\left[2\,n+l + \frac{1}{2}\left(3-\frac{d^2}{\ell^2}\right)\right]
\ ,
$
where $n$ is the radial quantum number and $l$ the angular momentum
(not to be confused with $\ell$).
Following the idea in Ref.~\cite{dvali12,dvali1207,dvali13,dvali13b}, we view the above spectrum as
representing the effective Quantum Mechanical dynamics of depleting modes, which can be described
by the first (non-rotating) excited state~\footnote{Note we have already integrated
out the angular coordinates.}
$
\psi_{100}(r)
=
\sqrt{\frac{2}{3\,\ell^7\,\sqrt{\pi}}}\,
e^{-\frac{r^2}{ 2\, \ell^2}}\, 
\left(2\,r^2-3\,\ell^2\right)
$.
The marginally binding condition~\eqref{energy0}, that is $E_{10}\simeq 0$, then
leads to the desired scaling laws
$
\ell
=
\sqrt{\frac{2\,N}{7}}\, \lp
\quad
{\rm and}
\quad
\mu
=
{\sqrt{\frac{7}{2\,N}}} \,\mpl
$.
We can now estimate the effect of the finite width of the potential
well~\eqref{Vho} by simply applying first order perturbation theory
and obtain
$
\Delta E_{10}
=
-\int_d^\infty r^2 \,\ud r \, \psi_{100}^2(r) \, V_0(r)
\simeq
-\frac{0.1}{\sqrt{N}}\,\mpl
\ .
$
This can now be compared, for example, with the ground state energy
$E_{00}=-\sqrt{14/N}\,\mpl\simeq -3.7\,\mpl/\sqrt{N}$.
Since $|\Delta E_{10}|\ll |E_{00}|$, our approximation appears reasonable.
We however remark that the ground state energy in this model has no physical meaning.
Indeed, the Schr\"odinger equation must be viewed as describing the
effective dynamics of BH constituents, and the total energy of the ``harmonic black hole'' is still given
by the sum of the individual boson effective masses,
\be
M=
N \,\mu
\simeq
\sqrt{\frac{7\,N}{2}}\,\mpl
\ ,
\label{Mtot}
\ee
in agreement with the ``maximal packing'' of Eq.~\eqref{Max} and the expected
mass spectrum of quantum BHs (see, for example, Refs.~\cite{bekenstein74,dvali1106}).

%
%

\section{Regular geometry}
\label{secReg}
It is now reasonable to assume that the actual density profile of the BEC gravitational source
is related to the ground state wave function in Eq.~\eqref{psin} according to
\be
\rho(r)
\simeq
M\,\psi_{000}^2
\simeq
\frac{7^2\,\mpl\,e^{-\frac{7\,r^2}{2\,N\,\lp^2}}}{\sqrt{\pi}\,N\,\lp^3}
\ .
\label{rhobec}
\ee
Similar Gaussians profiles have been extensively studied in
Refs.~\cite{nicolini09,nss06},
where it was proven that such densities satisfy the Einstein field equations
with a ``de~Sitter vacuum'' equation of state, $\rho = - p$,
where $p$ is the pressure.
Curiously, BECs can display this particular equation of state~\cite{chavanis11,chavanis12,Pitaevskii2009}.
This feature provides a connection between Quantum Mechanics and the geometrical description.
\par
Let us indeed take the static and normalised, energy density profile of
Ref.~\cite{nss06}\footnote{The squared length $\theta$ should not be confused with one of the angular coordinates of the previous expressions. Also, note $\rho$ has already been integrated over the angles.},
\be
\rho(r)
=
\frac{M\,e^{-\frac{r^2}{4\,\theta}}}{\sqrt{4\,\pi}\ \theta^{3/2}} \label{rhopiero}
\ ,
\ee
where $\sqrt{\theta}$ is viewed as a fundamental length
related to space-time noncommutativity, and $r$ is the radial coordinate
such that the integral inside a sphere of area $4\,\pi\,r^2$
gives the total Arnowitt-Deser-Misner (ADM) mass $M$ of the object for $r\to\infty$, i.e.:
$
M(r)
=
\int_0^r \rho(\bar r)\,\bar r^2\,\ud \bar r
=
M\, \frac{\gamma(3/2,r^2/4\theta)}{\Gamma(3/2)}
\ .
$
Here, $\Gamma(3/2)$ and $\gamma(3/2,r^2/4\theta)$ are the complete
and upper incomplete Euler Gamma functions, respectively.
This energy distribution then satisfies Einstein field equations together with the
Schwarzschild-like metric
$
\ud s^2
=
-f(r)\,\ud t^2
+ f^{-1}(r)\,\ud r^2
+ r^2\, \ud\Omega^2
\ ,
$
where
$
f(r)= 1 - \frac{2\,\gn\, M(r)}{r}
$.
According to Ref.~\cite{nss06}, one has a BH only if the
\emph{mass-to-characteristic length\/} ratio is sufficiently large,
namely for
\be
M
\gtrsim
1.9\,\frac{\sqrt{\theta}}{\gn}
=
1.9\,\mpl\,\frac{\sqrt{\theta}}{\lp}
\equiv
M_*
\ .
\label{Mth}
\ee
If the above inequality is satisfied, the metric function $f=f(r)$ has two zeros and
there are two distinct horizons.
For $M=M_*$, $f=f(r)$ has only one zero which corresponds to an ``extremal''
BH, with two coinciding horizons (and vanishing Hawking temperature).
The latter represents the minimum mass BH, and a candidate
BH remnant of the Hawking decay~\cite{cn08}.
Further, the classical Schwarzschild case is precisely recovered in the limit
$\gn\,M/\sqrt{\theta} \to\infty$, so that departures from the standard geometry
become quickly negligible for very massive BHs.
\par
Going back to the BEC model, whose total ADM mass is given in Eq.~\eqref{Mtot},
and comparing the Gaussian profile~\eqref{rhobec} with Eq.~\eqref{rhopiero},
that is setting $\theta=N\,\lp^2/14$,
one finds that the condition in Eq.~\eqref{Mth} reads
$
1.8\,\sqrt{N}
\gtrsim
0.5\,\sqrt{N}
\ ,
$
and is always satisfied (for $N\ge 1$).
We can therefore conclude that harmonic black holes always have two horizons,
and the degenerate case is not realised in their spectrum.
Although this mismatch might appear as a shortcoming of our construction,
it is actually consistent with the idea that the extremal case should have vanishing
Hawking temperature and therefore no depleting modes.
It also implies that the final evaporation phase, if it ends in the extremal case,
must be realised by a transition that most likely drives the BEC out of the critical point.
The precise nature of such a ``quantum black hole'' state remains, however,
unclear.
%
%
%
%
%
%
\section{Conclusions and outlook}
\label{secConc}
\setcounter{equation}{0}
We have shown that the scenario of Ref.~\cite{dvali12,dvali1207,dvali13,dvali13b}, in which BH inner degrees of freedom
(as well as the Hawking radiation) correspond to depleting states in a BEC, can be understood and
recovered in the context of General Relativity by viewing a BH as made of the superposition
of $N$ constituents, with a Gaussian density profile, whose characteristic length is given by the
 constituents' effective Compton wavelength.
From the point of view of Quantum Mechanics, such states straightforwardly arise from a binding
harmonic oscillator potential.
Moreover, requiring the existence of (at least) a horizon showed that the extremal case,
corresponding to a remnant with vanishing Hawking temperature, is not realised in the
harmonic spectrum~\eqref{Mtot}.
Such states will therefore have to be described by a different model. 
\par
At the threshold of BH formation (see, for example, Ref.~\cite{cmo12}),
for a total ADM mass $M\simeq \mpl$ (thus $N\simeq 1$),
the above description should allows us to describe Quantum Mechanical processes
involving BH intermediate (or metastable) states.
However, we can already anticipate that quantum BHs with spin should be
relatively easy to accommodate in our description, by simply considering
states in Eq.~\eqref{psin} with $l>0$.
This should allow us to consider more realistic quantum BH formation from particle
collisions.
parameter.
\par
Many questions are still left open.
First of all, the discretisation of the mass has an important consequence in the classical limit.
For example, let us look again at Eq.~\eqref{Mtot}, and consider two non-rotating BHs
with mass $M_1=\sqrt{\frac{7}{2}\,N_1} \, \mpl$ and $M_2=\sqrt{\frac{7}{2}\, N_2} \, \mpl$,
where $N_1$ and $N_2$ are positive integers, which slowly merge in a head-on collision
(with zero impact parameter).
The resulting BH should have a mass $M$ which is also given by Eq.~\eqref{Mtot}.
However, there is in general no integer $N_3$ such that
$\sqrt{N_3} = \sqrt{N_1} + \sqrt{N_2}$.
It therefore appears that either the mass should not be conserved, $M\not= M_1+M_2$,
or the mass spectrum described by Eq.~\eqref{Mtot} is not complete.  
This problem, which is manifestly more significant for small BH masses
(or, equivalently, integers $N$), is shared by all those models in which the the BH
mass does not scale exactly like an integer.
If we wish to keep Eq.~\eqref{Mtot}, or any equivalent mass spectrum, we might then argue
that a suitable amount of energy (of order $M_1+M_2-M_3$)
should be expelled during the merging, in order to accommodate the overall mass
into an allowed part of the spectrum.
In this case, one may also wonder if this emission can be thought of as
some sort of Hawking radiation~\footnote{Note that for vanishing impact parameter,
one does not expect any emission of classical gravitational waves.},
or if it is completely different in nature.
\par
Another issue regards the assumption in Eq.~\eqref{rhobec}, i.e.~the idea that the
classical density profile corresponds to the square modulus of the (normalised)
wavefunction.
At the semiclassical level, this seems reasonable and intuitive, but necessarily
removes the concept of ``point-like test particle'' from General Relativity,
thus forcing us to reconsider the idea of geodesics only in terms of propagation
of extended wave packets,
which might show unexpected features or remove others from the classical theory.
Also, elementary particles would not differ from extended massive objects and therefore
should have an equation of state (see, for instance, the old shell model in
Refs.~\cite{cgs09}).
\par
Last but not least, there is the question of describing the formation of a
BEC during a stellar collapse.
Condensation is usually achieved at extremely low temperature,
when the thermal de~Broglie wavelength becomes comparable to the
inter-particle spacing.
Whereas one has no doubt that particles inside a BH are extremely
packed, it is not clear how such a dramatic drop of temperature could occur.
%
%
\section*{Acknowledgements}
This work is supported in part by the European Cooperation
in Science and Technology (COST) action MP0905 ``Black Holes in a Violent  Universe". 
%
%
%

\begin{thebibliography}{99.}%
%
%
%
%
%
\bibitem{adm60}
R.~Arnowitt, S.~Deser and C.W.~Misner,
Phys. Rev. Lett. {\bf 4} (1960) 375;
%
%
\bibitem{bekenstein74}
J.D.~Bekenstein,
Lett.\ Nuovo Cim.\  {\bf 11} (1974) 467.
%
%
%
%
%
%
\bibitem{cn08} 
R.~Casadio and P.~Nicolini,
JHEP {\bf 0811}, 072 (2008).
%
%
\bibitem{cgs09}
R.~Casadio, R.~Garattini and F.~Scardigli,
Phys.\ Lett.\ B {\bf 679} (2009) 156.
%
%
\bibitem{cmo12} 
R.~Casadio, O.~Micu and A.~Orlandi,
Eur.\ Phys.\ J.\ C {\bf 72}, 2146 (2012).
%
\bibitem{chandra92}
S.~Chandrasekhar,
\textit{The mathematical theory of black holes},
Oxford University Press (1992).
%
%
\bibitem{chavanis11}
P.-H.~Chavanis,
Phys.\ Rev.\ D {\bf 84} (2011) 043531.
%
%
\bibitem{chavanis12}
P.-H.~Chavanis,
Astrophys.\ J.\  {\bf 537} (2012) A127;
%
%
\bibitem{dvali1005}
G.~Dvali and C.~Gomez,
\textit{Self-Completeness of Einstein Gravity}
arXiv:1005.3497 [hep-th].
%
%
\bibitem{dvali1106}
G.~Dvali, C.~Gomez and S.~Mukhanov,
\textit{Black Hole Masses are Quantized},
arXiv:1106.5894 [hep-ph].
%
%
\bibitem{dvali12}
G.~Dvali and C.~Gomez,
Phys.\ Lett.\ B {\bf 716} (2012) 240;
%
%
\bibitem{dvali1207} 
G.~Dvali and C.~Gomez,
\textit{Black Holes as Critical Point of Quantum Phase Transition},
arXiv:1207.4059 [hep-th];
%
%
\bibitem{dvali13}
G.~Dvali and C.~Gomez,
Fortsch.\ Phys.\  {\bf 61} (2013) 742.
%
%
\bibitem{dvali13b}
G.~Dvali and C.~Gomez,
Phys.\ Lett.\ B {\bf 719} (2013) 419;
%
%
\bibitem{hawking74} 
S.W.~Hawking,
Nature {\bf 248}, 30 (1974);
%
%
\bibitem{hawking75}
S.W.~Hawking,
Commun.\ Math.\ Phys.\  {\bf 43}, 199 (1975)
[Erratum-ibid.\  {\bf 46}, 206 (1976)].
%
%
\bibitem{nicolini09}
P.~Nicolini,
Int.\ J.\ Mod.\ Phys.\ A {\bf 24} (2009) 1229;
%
%
\bibitem{nss06}
P.~Nicolini, A.~Smailagic and E.~Spallucci,
Phys.\ Lett.\ B {\bf 632} (2006) 547.
%
%
\bibitem{nos11}
P.~Nicolini, A.~Orlandi and E.~Spallucci,
``The final stage of gravitationally collapsed thick matter layers,''
arXiv:1110.5332 [gr-qc].
%
%
\bibitem{Pitaevskii2009}
L.P.~Pitaevskii and S.~Stringari, \textit{Bose-Einstein condensation},
Oxford University Press (2003), Chapter~11

\end{thebibliography}
%

\end{document}